\documentclass[%
 reprint,
 amsmath,amssymb,
 aps,
prb,
]{revtex4-2}

\usepackage{graphicx}
\usepackage{dcolumn}
\usepackage{bm}
\usepackage{float}

\begin{document}


\title{Slope of $H_{c2}$ close to $T_c$ versus the size of the Cooper pairs: The role of disorder in Dynes superconductors}

\author{Franti\v{s}ek Herman and Richard Hlubina}

\affiliation{Department of Experimental Physics, Comenius University,
  Mlynsk\'{a} Dolina F2, 842 48 Bratislava, Slovakia}

\date{\today}

\begin{abstract}
The size of the Cooper pair $\xi_{\rm pair}$ is one of the basic characteristics of a superconductor, but it is not possible to measure it directly. It might be argued that $\xi_{\rm pair}$ can be determined from the value $\xi_{\rm slope}$ extracted from the measurement of the slope of $H_{c2}$ close to $T_c$. Taking into account both pair-conserving and pair-breaking scattering on impurities within the recently developed theory of Dynes superconductors, we perform an explicit calculation of $\xi_{\rm pair}$ and $\xi_{\rm slope}$. We show that the two quantities agree only in clean superconductors. In particular, when the pair-breaking disorder approaches the quantum critical point, $\xi_{\rm pair}$ stays finite, whereas $\xi_{\rm slope}$ diverges.
\end{abstract}
\maketitle

\section{Introduction}
A superconductor is characterized by three length scales: the size of the Cooper pairs $\xi_{\rm pair}$, the penetration depth $\lambda$ measuring the response of the condensate to a static applied magnetic field, and the typical distance $r_0$ between the electrons. While the ratio between $\xi_{\rm pair}$ and $\lambda$ is well known to determine whether the response of the superconductor is local or non-local \cite{Tinkham04}, the role of the ratio between $\xi_{\rm pair}$ and $r_0$ has been studied much less in the literature; see, however, Ref.~\cite{Strinati18} and references therein. 

In a typical superconductor the strong inequality $\xi_{\rm pair}\gg r_0$ is usually valid. In that case the Cooper pairs strongly overlap and the phase fluctuations may be expected to be small. This is the situation considered within the standard BCS theory. In the opposite extreme case, $\xi_{\rm pair}\ll r_0$, the Cooper pairs form well-defined bosons and the symmetry-breaking transition towards the superconducting state can be understood within the framework of the Bose-Einstein condensation (BEC) \cite{Nozieres85}. Therefore, in order to distinguish between the role played by the BCS and BEC mechanisms in a given material, it is important to know the size of the Cooper pairs $\xi_{\rm pair}$. 

Unfortunately, $\xi_{\rm pair}$ is not directly accessible experimentally. In order to circumvent this complication, in a recent paper addressing this issue in the high-temperature superconductors, it has been suggested that the position in the BCS-BEC spectrum can be determined from a new length scale $\xi_{\rm slope}$ \cite{Chen23}. Namely, it was suggested to  measure the slope of the upper critical field $B_{c2}$ close to the critical temperature $T_c$ and to determine $\xi_{\rm slope}$ from the following equation:
\begin{equation}
\left.\frac{\partial B_{c2}}{\partial T}\right |_{T=T_c}=-\frac{\Phi_0}{2\pi T_c \xi_{\rm slope}^2},
\label{eq:xi_slope}
\end{equation}
where $\Phi_0$ is the superconducting flux quantum. Since, according to standard understanding \cite{Nozieres85}, it is the ratio $\xi_{\rm pair}/r_0$ which determines the position in the BCS-BEC spectrum, one might expect that $\xi_{\rm slope}$ represents an easily accessible experimental proxy for $\xi_{\rm pair}$.

The goal of this paper is to decide whether $\xi_{\rm slope}$ does indeed provide a reasonable estimate of $\xi_{\rm pair}$, at least in the BCS-like limit $\xi_{\rm pair}\gg r_0$. To this end, we will make use of the recently developed theory of Dynes superconductors, which can be viewed as a minimal extension of the BCS theory taking into account the presence of disorder \cite{Herman16}. 

Depending on their action on the Cooper pairs, in a superconductor there exist two types of impurities: pair-conserving or pair-breaking. Within the theory of Dynes superconductors \cite{Herman16}, pair-conserving scattering on a random scalar field and pair-breaking scattering on a random magnetic field are considered. Both types of fields are assumed to be spatially uncorrelated and their action on the superconducting state is treated within the coherent potential approximation \cite{Elliott74}.  The central result of the theory is that, provided the distribution of magnetic fields is described by a Lorentzian and the distribution of scalar fields is even but otherwise arbitrary, a simple analytical formula for the electrons' Green function can be written down \cite{Herman16}.

Previously we have shown that the matrix Green's function $\hat{G}$ of a Dynes superconductor has several favourable properties \cite{Herman17a}: it is analytic in the upper half-plane, it has the correct large-frequency asymptotics, its diagonal spectral functions are positive definite, and it satisfies the sum rules for the zero-order moment of the spectral function. Therefore, we believe, $\hat{G}$ can be used as a generic two-lifetime Green's function of a superconductor.

The plan of this paper is as follows. In Section~2 we start by calculating the anomalous spectral function of a Dynes superconductor. We will also show how to calculate the superconducting order parameter as well as the internal wavefunction of the Cooper pair. 

In Section~3, we present a direct calculation of the pair size $\xi_{\rm pair}$ within the Dynes theory. We will study in detail the dependence of $\xi_{\rm pair}$ on the parameters characterizing the Dynes superconductor: gap size $\Delta$, pair-conserving scattering rate $\Gamma_s$, pair-breaking scattering rate $\Gamma$, and temperature $T$. In particular, we want to decide whether, in the limit when $\Gamma$ approaches the quantum critical point where superconductivity disappears, the pair size $\xi_{\rm pair}$ diverges (as one might naively expect since $\Delta\rightarrow 0$ in this limit) or not.

In Section~4 we present the results for the length scale $\xi_{\rm slope}$, which can be simply obtained from the Ginzburg-Landau analysis of the Dynes superconductors already presented in Ref.~\cite{Herman18}. Next, we will compare the results for $\xi_{\rm pair}$ and $\xi_{\rm slope}$. We will show that, in the textbook case of a clean BCS-like superconductor, the two quantities do in fact agree as assumed in Ref.~\cite{Chen23}, up to a trivial difference in normalization. However, in presence of impurities there exist important qualitative differences between $\xi_{\rm pair}$ with $\xi_{\rm slope}$. In particular, we will show that the difference between $\xi_{\rm pair}$ and its proxy is largest in presence of strong pair-breaking scattering. 

Finally, in Section~5 we will present our conclusions.

\section{Anomalous propagator}
We consider a single band of electrons in an isotropic singlet pairing state. Within the Nambu-Gor'kov formalism, the Green's function of the superconductor $\hat{G}({\bf k},\omega)$ is a $2\times 2$ matrix. Therefore it can be written as a sum of components proportional to the $2\times 2$ unit matrix, $\tau_0$, and to the Pauli matrices $\tau_i$ with $i=1,\ldots,3$. The diagonal components of $\hat{G}({\bf k},\omega)$ describe the propagation of electrons and holes with momentum ${\bf k}$ and energy $\omega$, while the off-diagonal components correspond to the so-called anomalous propagator. 

Within the theory of Dynes superconductors, the anomalous component '12' of the Green's function of a superconductor with gap $\Delta$ in presence of the pair-conserving scattering rate $\Gamma_s$ and pair-breaking scattering rate $\Gamma$ is given by the expression \cite{Herman16}
\begin{equation}
\hat{G}_{12}(\varepsilon_{\bf k},\omega)=
\frac{\Delta}{2\Omega}
\left(\frac{1}{\Omega+i\Gamma_s-\varepsilon_{\bf k}}
+\frac{1}{\Omega+i\Gamma_s+\varepsilon_{\bf k}}\right),
\label{eq:g12}
\end{equation}
where the $\omega$-dependent energy scale $\Omega$ is given by
\begin{equation}
\Omega(\omega)=\sqrt{(\omega+i\Gamma)^2-\Delta^2}\equiv \Omega_1+i\Omega_2.
\label{eq:dynes_omega}
\end{equation}
Here we take that branch of the square root which has the property that the imaginary part of the root of a complex number is positive. Denoting the real and imaginary parts of $\Omega$ as $\Omega_1$ and $\Omega_2$, respectively, this sign convention implies that $\Omega_1(\omega)$ is an odd function of $\omega$, while $\Omega_2(\omega)$ is even.

For future convenience let us note that the Green's function depends on the momentum ${\bf k}$ only via the single-particle energy $\varepsilon_{\bf k}$, and therefore $\hat{G}_{12}({\bf k},\omega)$ is effectively equal to $\hat{G}_{12}(\varepsilon_{\bf k},\omega)$.

In what follows we adopt the following notation. We denote the gap of a system without pair breaking (i.e. for $\Gamma=0$) at temperature $T=0$ as $\Delta_{00}$. Under $\Delta(0)$ we understand the gap of a system with finite pair breaking at $T=0$; we have shown that $\Delta(0)^2=\Delta_{00}(\Delta_{00}-2\Gamma)$ \cite{Herman16}. The symbol $\Delta$ without indices is reserved for the gap of a superconductor with a general set of parameters $\Gamma$ and $T$. 

Similarly, $T_{c0}$ denotes the critical temperature in a system with $\Gamma=0$, while $T_c$ is the critical temperature for a finite $\Gamma$. It can be shown that for $\Gamma$ close to the maximal admissible value of $\Delta_{00}/2$, we have $(\pi T_c)^2=3\Delta_{00}(\Delta_{00}-2\Gamma)/2$. 

For the sake of completeness we remind the reader that pair-conserving scattering $\Gamma_s$ does not influence the values of $\Delta$, $\Delta(0)$, and $T_c$, in agreement with the Anderson theorem \cite{Anderson59}.

\subsection{Anomalous spectral function}

The spectral function of the anomalous propagator is given by $A_{12}(\varepsilon_{\bf k},\omega) \equiv -\pi^{-1} {\rm Im}\left[\hat{G}_{12}(\varepsilon_{\bf k},\omega)\right]$. Plugging Eq.~\eqref{eq:g12} into this definition, after some work one can find the following expression for the spectral function,
\begin{eqnarray}
  A_{12}(\varepsilon_{\bf k},\omega) &=&
  P\left[\delta_{\tilde{\Gamma}}(\Omega_1-\varepsilon_{\bf k})
    +\delta_{\tilde{\Gamma}}(\Omega_1+\varepsilon_{\bf k})\right]
  \nonumber
  \\
  &&+Q\delta_{\tilde{\Gamma}}(\Omega_1-\varepsilon_{\bf k})
  \delta_{\tilde{\Gamma}}(\Omega_1+\varepsilon_{\bf k}),
\label{eq:spectral_f}
\end{eqnarray}
where we have introduced $\tilde{\Gamma}=\Gamma_s+\Omega_2$ and the symbol $\delta_{\tilde{\Gamma}}(x)=\pi^{-1}\tilde{\Gamma}/(x^2+\tilde{\Gamma}^2)$ denotes a Lorentzian with width $\tilde{\Gamma}$. The $\omega$-dependent weights $P$ and $Q$  are given by
\begin{eqnarray*}
  P=\frac{\Delta \Omega_1}{2(\Omega_1^2+\Omega_2^2)}
  \frac{\Gamma_s}{\tilde{\Gamma}},
  \quad
Q=\frac{2\pi \Delta \Omega_1\Omega_2}{\Omega_1^2+\Omega_2^2}
  \frac{\Omega_1^2+\tilde{\Gamma}^2}{\tilde{\Gamma}^2}.
\end{eqnarray*}
One finds readily that $A_{12}(\varepsilon_{\bf k},\omega)$ exhibits the following symmetries:
\begin{align*}
A_{12}(\varepsilon_{\bf k},-\omega)= - A_{12}(\varepsilon_{\bf k},\omega),
\quad
A_{12}(-\varepsilon_{\bf k}, \omega)= A_{12}(\varepsilon_{\bf k}, \omega).
\end{align*}

\begin{figure*}[t]
\includegraphics[width = 17cm]{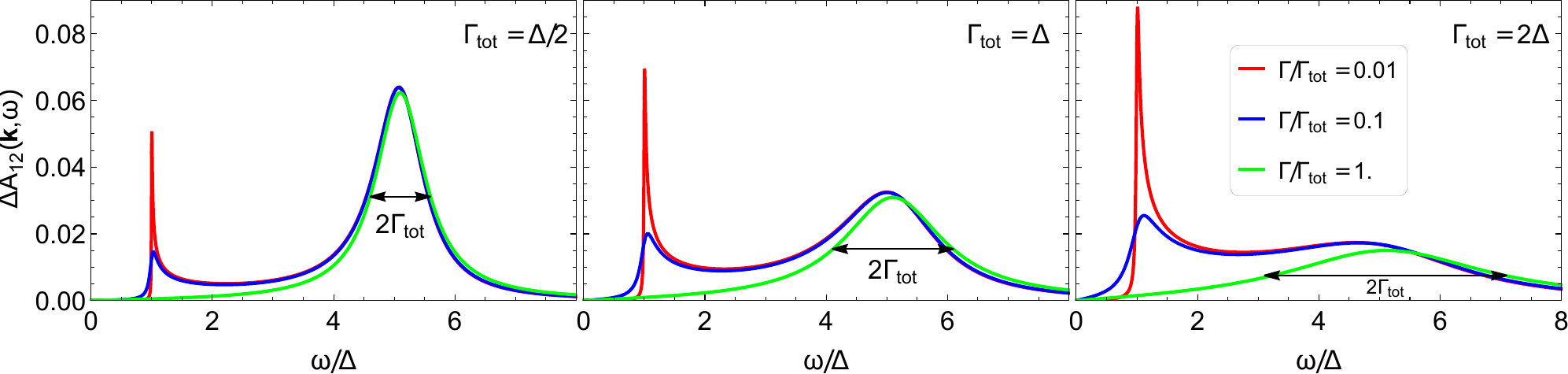}
\caption{Anomalous spectral functions $A_{12}(\mathbf{k},\omega)$ of the Dynes superconductor for energy $\varepsilon_{\bf k}=5\Delta$. The total scattering rate $\Gamma_{\rm tot}=\Gamma+\Gamma_s$ increases from the left to the right panel. The curves in each panel differ by the strength of the pair-breaking scattering rate $\Gamma$, while $\Gamma_{\rm tot}$ is kept fixed. The color coding is the same in all panels.}
\label{fig:spect_fun}
\end{figure*}

Setting the pair-conserving rate to $\Gamma_s=0$, the spectral function Eq.~\eqref{eq:spectral_f} simplifies to
$$
A_{12}(\varepsilon_{\bf k},\omega)=\frac{\Delta}{2E_{\bf k}}
\left[\delta_\Gamma(\omega-E_{\bf k})-\delta_\Gamma(\omega+E_{\bf k})\right],
$$
where $E_{\bf k}=\sqrt{\varepsilon_{\bf k}^2+\Delta^2}$ is the quasiparticle energy in the superconducting state. As a function of $\omega$, the spectral function is thus seen to be the difference of two Lorentzians at $\pm E_{\bf k}$, which reduce to delta-functions in the BCS case where $\Gamma=0$.

Setting the pair-breaking rate to $\Gamma=0$, we find that the anomalous spectral function is non-vanishing only for $|\omega|>\Delta$. For positive $\omega$ we find 
$$
A_{12}(\varepsilon_{\bf k},\omega)=\frac{\Delta}{2\Omega}
\left[\delta_{\Gamma_s}(\Omega-\varepsilon_{\bf k})
+\delta_{\Gamma_s}(\Omega+\varepsilon_{\bf k})\right],
$$
where $\Omega=\sqrt{\omega^2-\Delta^2}$. Note that the Lorentzians are peaked at the same energies $\omega=\pm E_{\bf k}$ as in the BCS case. However, finite pair-conserving scattering generates also two new (divergent) peaks of $A_{12}(\varepsilon_{\bf k},\omega)$ at energies $\pm\Delta$. 

Numerical results for $A_{12}(\varepsilon_{\bf k},\omega)$ when both types of scattering are present show that the peaks at $\pm E_{\bf k}$ acquire a finite width, roughly given by the total scattering rate $\Gamma_{\rm tot}=\Gamma+\Gamma_s$. Also the peaks at $\pm \Delta$ are smeared by a finite value of the pair-breaking scattering rate $\Gamma$. These results are very similar to those for the diagonal spectral function $A_{11}(\varepsilon_{\bf k},\omega)$ obtained in Ref.~\cite{Herman17a}. For an explicit example, see Fig.~\ref{fig:spect_fun}. There we plot $A_{12}(\varepsilon_{\bf k},\omega)$ only for $\omega>0$, since it is an odd function of frequency $\omega$. It is worth pointing out that for momenta at the Fermi surface, i.e. for $\varepsilon_{\bf k}=0$, the two peaks of $A_{12}(\varepsilon_{\bf k},\omega)$ at positive $\omega$ merge into a single one at $\omega\approx \Delta$.

\subsection{Cooper pair wavefunction}

The internal wavefunction of the Cooper pair $\varphi({\bf r})$ depends on the relative distance between the electrons forming the pair. It is given by $\varphi({\bf r})\equiv\langle \psi_\uparrow({\bf x}) \psi_\downarrow({\bf x}+{\bf r})\rangle$, where $\psi_\sigma({\bf x})$ annihilates an electron with spin $\sigma$ at lattice site ${\bf x}$. Fourier transforming and introducing annihilation operators $c_{{\bf k}\sigma}$ for electrons in Bloch states with momentum ${\bf k}$ and spin $\sigma$, we thus obtain
\begin{equation}
\varphi({\bf r})=\frac{1}{\cal N}\sum_{\bf k}b_{\bf k} e^{-i{\bf k}\cdot{\bf r}},
\label{eq:wavefunction}
\end{equation}
where we have introduced the superconducting order parameter $b_{\bf k} \equiv \langle c_{{\bf k}\uparrow} c_{{-\bf k}\downarrow}\rangle $ and ${\cal N}$ is the number of lattice sites.

In Ref.~\cite{Herman17a} it has been noted that, quite generally, the order parameter $b_{\bf k}$ is related to the anomalous spectral function by a sum rule. Exploiting the fact that $A_{12}(\varepsilon_{\bf k},\omega)$ is an odd function of $\omega$, the sum rule (B4) in \cite{Herman17a} simplifies to the following expression for the order-parameter function $b(\varepsilon_{\bf k})$:
\begin{equation}
b_{\bf k}=b(\varepsilon_{\bf k}) = 
\int_0^\infty d\omega A_{12}(\varepsilon_{\bf k},\omega)\tanh{\frac{\omega}{2T}}.
\label{eq:bk}
\end{equation}
Making use of Eqs.~(\ref{eq:spectral_f},\ref{eq:wavefunction},\ref{eq:bk}), one can in principle calculate the full wavefunction of the Cooper pair.

Turning to the order parameter function $b(\varepsilon_{\bf k})$, let us start by quoting the well-known result for a clean BCS superconductor, $b(\varepsilon_{\bf k})=\frac{\Delta}{2E_{\bf k}}\tanh \frac{E_{\bf k}}{2T}$. The function $b(\varepsilon_{\bf k})$ is even, with a maximum at the Fermi level $\varepsilon_{\bf k}=0$. The value of $b(0)$ decreases from $b(0)=1/2$ at $T=0$ to $b(0)=\Delta/(4T_{c0})$ close to the critical temperature. As a function of $|\varepsilon_{\bf k}|$, the order parameter decreases, varying ultimately at large $|\varepsilon_{\bf k}|$ as $b\approx \Delta/(2|\varepsilon_{\bf k}|)$. The function is appreciable for $|\varepsilon_{\bf k}|\lesssim \Delta_{00}$ at $T=0$ and for $|\varepsilon_{\bf k}|\lesssim 2T_{c0}$ close to the critical temperature.

The function $b(\varepsilon_{\bf k})$ exhibits qualitatively similar behavior also for finite $\Gamma$ and $\Gamma_s$. For instance, if $\Gamma_s=0$ and $T=0$, we find $b(\varepsilon_{\bf k})=\tfrac{\Delta}{\pi E_{\bf k}}\arctan \tfrac{E_{\bf k}}{\Gamma}$. Thus, with changing parameters, the shape of $b(\varepsilon_{\bf k})$ can be roughly parameterized by two parameters: the height of the maximum, $b(0)$, and the width of the maximum in $\varepsilon_{\bf k}$-space. It is the width of the maximum which will turn out to be relevant for determination of the pair size, see Eq.~\eqref{eq:xi} in the next Section.

\section{Size of the Cooper pair}
Following standard procedures \cite{DeGennes66}, once the wavefunction $\varphi({\bf r})$ of the Cooper pair is known, we define the size of the pair $\xi_{\rm pair}$ as the mean square distance between the electrons forming the pair,
\begin{equation*}
\xi_{\rm pair}^2 \equiv \frac{\int dV r^2 |\varphi({\bf r})|^2}{\int dV |\varphi({\bf r})|^2} = \frac{\sum_{\bf k} \left(\frac{\partial b_{\bf k}}{\partial {\bf k}}\right)^2}{\sum_{\bf k} b_{\bf k}^2}.
\end{equation*}
The second expression follows from the fact that both $\varphi({\bf r})$ and $b_{\bf k}$ are even. 

If we take into account that $b_{\bf k}$ depends on the momentum ${\bf k}$ only via the quasiparticle energy $\varepsilon_{\bf k}$, we have $\partial b_{\bf k}/\partial {\bf k}=\hbar {\bf v}_{\bf k}\partial b/\partial \varepsilon_{\bf k}$ where ${\bf v}_{\bf k}$ is the quasiparticle velocity. Approximating the velocity of all quasiparticles by the Fermi velocity $v_F$ we therefore finally find that the size of the Cooper pair is given by the expression
\begin{equation}
\xi_{\rm pair}= \frac{\hbar v_F}{\Lambda},
\qquad
\frac{1}{\Lambda^2}\equiv
\frac{\int_0^\infty d\varepsilon_{\bf k} \left(\frac{\partial b(\varepsilon_{\bf k})}{\partial \varepsilon_{\bf k}}\right)^2}{\int_0^\infty d\varepsilon_{\bf k} b(\varepsilon_{\bf k})^2}.
\label{eq:xi}
\end{equation}

As explained in the Appendix, the integrals entering the fraction defining the energy scale $\Lambda$ in Eq.~\eqref{eq:xi} can be alternatively calculated also on the imaginary axis. This latter formulation is especially useful for numerical calculations. 

In a Dynes superconductor, the energy scale $\Lambda$ depends on the parameters entering Eq.~\eqref{eq:g12}, i.e. $\Delta$, $\Gamma_s$, and $\Gamma$, as well as on the temperature $T$. Let us start by presenting the results for $\xi_{\rm pair}$ in several special cases.

a) {\it BCS case without impurities: $\Gamma_s=0$ and $\Gamma=0$.} At temperature $T=0$, taking the integrals in Eq.~\eqref{eq:xi} we find that the energy scale $\Lambda=\sqrt{8}\Delta_{00}\approx 4.99 T_{c0}$, the latter equality following from the BCS ratio $\Delta_{00}/T_{c0}\approx 1.764$. On the other hand, close to the critical temperature $T\rightarrow T_{c0}$ we similarly find $\Lambda\approx 6.10 T_{c0}$. This means that, as is well known, the energy scale $\Lambda$ changes only mildly between $T=0$ and $T=T_{c0}$. Note that, somewhat surprisingly, the Cooper pair is slightly smaller at higher temperatures. For future considerations it is important to point out that the finite value of $\Lambda$ is rendered by the finite value of $\Delta$ at $T=0$,
and by the finite value of temperature at $T=T_c$.

b) {\it Pair-breaking rate approaching the quantum critical point, $\Gamma\rightarrow \Delta_{00}/2$, but $\Gamma_s=0$.} In this case, since both $\Delta(0)$ and $T_c$ are small, one can write $b(\varepsilon_{\bf k})\approx \tfrac{\Delta}{\pi\varepsilon_{\bf k}}\arctan\tfrac{\varepsilon_{\bf k}}{\Gamma}$, and taking the integrals in Eq.~\eqref{eq:xi} leads to $\Lambda\approx 3.91 \Gamma \approx 1.95 \Delta_{00}\approx 3.45 T_{c0}$. Note that, compared with the pure BCS case at $T=0$, the energy scale $\Lambda$ exhibits only a minor decrease. This is one of the main results of this paper. We stress that the finite value of $\Lambda$ is a very surprising result, since naively one might expect that the energy $\Lambda$ scales with $\Delta(0)$, which vanishes at the critical point! The finite value of $\Lambda$ is rendered by the finite value of $\Gamma$, as can be observed, e.g., from the shape of the function $b(\varepsilon_{\bf k})$.

\begin{figure}[t!]
\includegraphics[width = 7.5 cm]{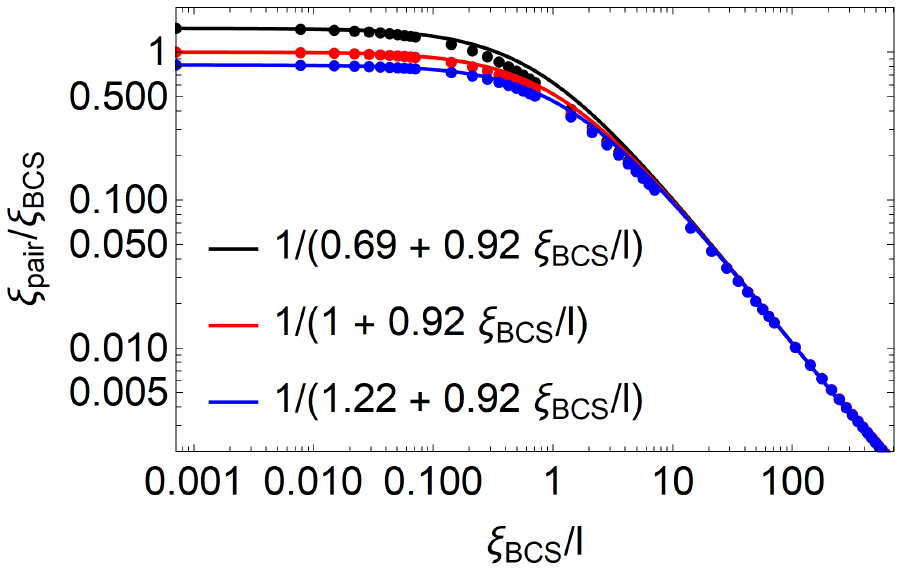}
\caption{Cooper pair size $\xi_{\rm pair}$ in units of $\xi_{\rm BCS}$ as a function of the pair-conserving rate $\Gamma_s$, parameterized as $\xi_{\rm BCS}/\ell$. Black symbols: pair-breaking rate $\Gamma\rightarrow \Delta_{00}/2$ and temperature $T=0$. Red symbols: $\Gamma=0$ and $T= 0$. Blue symbols: $\Gamma=0$ and $T\rightarrow T_c$.}
\label{fig:dirty}
\end{figure}

c) {\it Finite value of $\Gamma_s$ in absence of pair-breaking scattering, $\Gamma=0$.} This case is considered numerically in Fig.~\ref{fig:dirty}. The pair size is measured in units of $\xi_{\rm BCS}\equiv \hbar v_F/(\sqrt{8}\Delta_{00})$, which is the pair size of a clean BCS superconductor at $T=0$. The strength of the pair-conserving scattering $\Gamma_s$ is replaced by the more commonly used mean free path $\ell\equiv \hbar v_F/(2\Gamma_s)$ \cite{note}. One can observe that, both at $T=0$ and at $T\rightarrow T_c$, in the studied range of mean free paths $\ell$ the pair size is reasonably described by a Pippard-like formula \cite{limit}
\begin{equation}
\frac{1}{\xi_{\rm pair}}=\frac{a}{\xi_{\rm BCS}}+\frac{b}{\ell}.
\label{eq:pippard}
\end{equation}
The numerical coefficient $a$ is equal to $a=1$ for vanishing temperature $T$, and $a=1.22$ for $T$ close to $T_c$, in agreement with the results in case a). At both temperatures, we find that $b=0.92$.

d) {\it Finite value of $\Gamma_s$ and nearly critical pair breaking, $\Gamma\rightarrow \Delta_{00}/2$.} As shown in Fig.~\ref{fig:dirty}, also in this case the pair size is reasonably described by the Pippard-like formula Eq.~\eqref{eq:pippard} with the same coefficient $b=0.92$. For the coefficient $a$ we find $a=0.69$ in agreement with the results in case b).

\begin{figure}[t!]
\includegraphics[width = 7.5 cm]{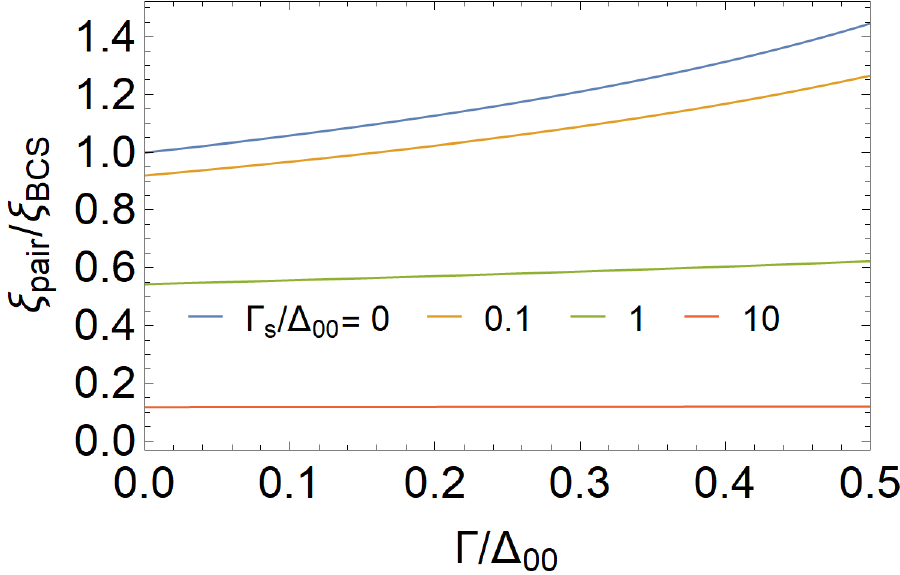}
\caption{Cooper pair size $\xi_{\rm pair}$ in units of $\xi_{\rm BCS}$ at $T = 0$ as a function of the pair-breaking rate $\Gamma$ in the full admissible range of $\Gamma$. Top to bottom curves correspond to $\Gamma_s/\Delta_{00}=0$, 0.1, 1, and 10, respectively. Note that $\xi_{\rm pair}$ is finite for all values of $\Gamma$.}
\label{fig:xi_vs_gamma}
\end{figure}

Having established how the pair size $\xi_{\rm pair}$ scales with the pair-conserving rate $\Gamma_s$, in Fig.~\ref{fig:xi_vs_gamma} we show the dependence of $\xi_{\rm pair}$ at temperature $T = 0$ on the pair-breaking rate $\Gamma$ in the full admissible range of $\Gamma$. Note that $\xi_{\rm pair}$ is finite for all values of $\Gamma$.

Taken together, the results for $\xi_{\rm pair}$ obtained in the various special cases lead us to conclude that the order of magnitude of the energy scale $\Lambda$ is given by $\Lambda\sim {\rm max}(\Delta, T, \Gamma, \Gamma_s)$. The crucial point to observe is that, when at least one of the energy scales $\Delta$, $T$, $\Gamma$, and $\Gamma_s$ is non-vanishing, also the pair size $\xi_{\rm pair}$ is finite.

\section{Comparison between $\xi_{\rm pair}$ and $\xi_{\rm slope}$}
Let us start by observing that, in the vicinity of the critical temperature $T_c$ \cite{note2}, the upper critical field is given by the expression $B_{c2}=\Phi_0/(2\pi\xi_{\rm GL}^2)$, where $\xi_{\rm GL}$ is the Ginzburg-Landau coherence length, which is known to diverge as $\xi_{\rm GL}={\rm const}/(1-t)^{1/2}$, where $t=T/T_c$ \cite{Tinkham04}. If we plug this expression into the definition~\eqref{eq:xi_slope}, we observe that there exists a simple relation between $\xi_{\rm GL}$ and $\xi_{\rm slope}$, namely
\begin{equation}
\xi_{\rm GL}=\frac{\xi_{\rm slope}}{\sqrt{1-t}}.
\label{eq:GL_vs_slope}
\end{equation}
Since the Ginzburg-Landau coherence length has already been calculated within the theory Dynes superconductors \cite{Herman18}, the results for $\xi_{\rm slope}$ can be found readily.  

For convenience, we will start by discussing the same four cases a) to d) as in the previous Section. In order to keep contact with the literature, in what follows we introduce the usual definition of the Pippard coherence length, $\xi_0\equiv \hbar v_F/(\pi \Delta_{00})$, and we note that it differs from $\xi_{\rm BCS}$ only by a numerical factor of order 1, $\xi_0\approx 0.9\xi_{\rm BCS}$.

a) $\Gamma_s=0$, $\Gamma=0$. In this clean BCS case we find
\begin{equation}
\xi_{\rm slope}\approx 0.74\xi_0 \approx 0.66\xi_{\rm BCS},
\end{equation}
in agreement with the literature \cite{Tinkham04}. This should be compared with the actual pair size $\xi_{\rm pair}=\xi_{\rm BCS}$ at $T=0$, or with $\xi_{\rm pair}\approx 0.82\xi_{\rm BCS}$ close to $T_c$. One observes that, up to a minor difference in the normalization, the size of the Cooper pair can in fact be determined from $\xi_{\rm slope}$ in the clean BCS case. This is of course well known.

b) $\Gamma_s=0$, $\Gamma\rightarrow \Delta_{00}/2$. In this somewhat academic case the findings of \cite{Herman18} imply that
\begin{equation}
\xi_{\rm slope}=
\frac{\pi}{\sqrt{12}}\frac{\xi_0}{\sqrt{1-2\Gamma/\Delta_{00}}}
\approx 0.82 \frac{\xi_{\rm BCS}}{\sqrt{1-2\Gamma/\Delta_{00}}}.
\end{equation}
This means that, when the quantum critical point at $\Gamma=\Delta_{00}/2$ is approached, the estimated pair size $\xi_{\rm slope}$ diverges. However, in the previous Section we have shown that the actual pair size $\xi_{\rm pair}$ stays finite in this limit. Thus, in presence of finite pair-breaking scattering, $\xi_{\rm slope}$ can not be used as a proxy for $\xi_{\rm pair}$.

c) $\Gamma_s\gg \Delta_{00}$, $\Gamma=0$. Making use of the results in Ref.~\cite{Herman18} we find that, in a dirty superconductor in absence of pair-breaking processes,
\begin{equation}
\xi_{\rm slope}\approx 0.85\sqrt{\xi_0 \ell},
\end{equation}
in agreement with literature \cite{Tinkham04}. This result should be compared with the Pippard-like expression Eq.~\eqref{eq:pippard} in the dirty limit, according to which  $\xi_{\rm pair}=\ell/b$. Thus $\xi_{\rm slope}$ and $\xi_{\rm pair}$ scale with different powers of the mean free path $\ell$, indicating that $\xi_{\rm slope}$ can not be a reasonable proxy for $\xi_{\rm pair}$ in the dirty limit.

d) $\Gamma_s\gg \Delta_{00}$, $\Gamma\rightarrow \Delta_{00}/2$. In a dirty superconductor with pair-breaking processes which nearly destroy the superconducting state we find
\begin{equation}
\xi_{\rm slope}=
\sqrt{\frac{\pi}{6}}\sqrt{\frac{\xi_0 \ell}{1-2\Gamma/\Delta_{00}}}.
\end{equation}
Similarly as in case b), the proxy $\xi_{\rm slope}$ diverges upon approaching the quantum critical point for disappearance of superconductivity, whereas the actual pair size stays finite. 

\section{Conclusions}
In conclusion, we have shown that, in a BCS-like isotropic ($s$-wave) superconductor described by the Dynes phenomenology \cite{Herman16}, the experimentally accessible length scale $\xi_{\rm slope}$ can serve as a proxy for the actual Cooper pair size $\xi_{\rm pair}$ only provided impurity scattering can be neglected.

When at least one of the scattering rates $\Gamma_s$ and $\Gamma$ can not be neglected, the length scales $\xi_{\rm slope}$ and $\xi_{\rm pair}$ are different. The difference between them is most spectacular if the quantum critical point of the superconductor to normal metal transition is approached by increasing pair-breaking scattering: The quantity $\xi_{\rm slope}$ diverges in this limit, but the actual pair size $\xi_{\rm pair}$ stays finite. 

The finite value of $\xi_{\rm pair}$ at the quantum critical point follows from the following argument. For dimensional reasons, $\xi_{\rm pair}$  is given by $\xi_{\rm pair}=\hbar v_F/\Lambda$, where $\Lambda$ is an appropriate energy scale. Although the energy scales $T$ and $\Delta$ both vanish at the quantum critical point, the pair-breaking scattering rate $\Gamma$ is necessarily finite, implying a finite value of $\Lambda$.

On the other hand, for superconductors in the dirty limit $\ell\ll\xi_0$ with small pair-breaking scattering, the situation changes completely: in this case we find $\xi_{\rm pair}\ll\xi_{\rm slope}$. Thus, even within the simple case of an isotropic BCS-like superconductor with impurities, the ratio $\xi_{\rm slope}/\xi_{\rm pair}$ can take any value from very small to very large ones.

The paper \cite{Chen23}, which motivated our discussion here, deals primarily with the cuprates. It should be stressed that our results do not directly apply to these materials, at the very least because of the $d$-wave symmetry of their pairing state. However, we do not see any arguments why, in the case of cuprates, the relation between $\xi_{\rm slope}$ and $\xi_{\rm pair}$ should turn to a simple one.  

\appendix
\section{Evaluation of $\xi_{\rm pair}$ on the Matsubara axis}
When analytically continued to the imaginary axis, Eq.~\eqref{eq:g12} can be written as
\begin{equation*}
{\hat G}_{12}(\varepsilon_{\bf k},\omega_n)=
-\frac{\Delta(1+\Gamma_s/\Omega_n)}{(\Omega_n+\Gamma_s)^2+\varepsilon_{\bf k}^2},
\end{equation*}
where we have defined $\Omega_n=\sqrt{(|\omega_n|+\Gamma)^2+\Delta^2}$, and $\omega_n=(2n+1)\pi T$ is the fermionic Matsubara frequency.

The key point to observe is that $b_{\bf k}$ can be written in terms of the Matsubara Green's function in the imaginary time $\tau=0_{+}$ approaching 0 from the right, $b_{\bf k}=b(\varepsilon_{\bf k})=-{\hat G}_{12}(\varepsilon_{\bf k},\tau=0_{+})=-T\sum_n {\hat G}_{12}(\varepsilon_{\bf k},\omega_n)$, see e.g. \cite{Rickayzen80}. Noticing furthermore that $\Omega_n$ is an even function of $\omega_n$, we thus find
$$
b(\varepsilon)=2\Delta T \sum_{\omega_n>0}^\infty 
\frac{1+\Gamma_s/\Omega_n}{(\Omega_n+\Gamma_s)^2+\varepsilon^2}.
$$
With this formula for $b(\varepsilon)$, the integrals over $\varepsilon$ entering Eq.~\eqref{eq:xi} are elementary and we find
\begin{eqnarray*}
\int_0^\infty d\varepsilon b^2(\varepsilon)&=&
\frac{\pi}{2}\sum_{n=0}^\infty \sum_{m=0}^\infty 
\frac{(2\Delta T)^2}{\Omega_n\Omega_m(\Omega_n+\Omega_m+2\Gamma_s)},
\nonumber
\\
\int_0^\infty d\varepsilon (\partial b/\partial \varepsilon)^2&=&
\pi\sum_{n=0}^\infty \sum_{m=0}^\infty 
\frac{(2\Delta T)^2}{\Omega_n\Omega_m(\Omega_n+\Omega_m+2\Gamma_s)^3}.
\end{eqnarray*}
These expressions are suitable for a fast numerical evaluation of the energy scale $\Lambda$ and, via Eq.~\eqref{eq:xi}, of the Cooper pair size $\xi_{\rm pair}$.

\begin{acknowledgments}
This work was supported by the Slovak Research and Development Agency under Contract No.~APVV-19-0371 and by the European Union’s Horizon 2020 research and innovation programme under the Marie Sklodowska-Curie Grant Agreement No. 945478.
\end{acknowledgments}


\begin{thebibliography}{99}
\bibitem{Tinkham04} M. Tinkham, {\it Introduction to Superconductivity}, 2nd ed. (Dover, New York, 2004).

\bibitem{Strinati18} G. C. Strinati, P. Pieri, G. R\"{o}pke, P. Schuck, and M. Urban, Phys. Rep. {\bf 738}, 1 (2018).

\bibitem{Nozieres85}P. Nozières and S. Schmitt-Rink, J. Low Temp. Phys. 59, 195 (1985).

\bibitem{Chen23} Q. Chen, Z. Wang, R. Boyack, and K. Levin, arXiv:2307.08611.

\bibitem{Herman16} F. Herman and R. Hlubina, Phys. Rev. B {\bf 94}, 144508 (2016).

\bibitem{Elliott74} R. J. Elliott, J. A. Krumhansl, and P. L. Leath, Rev. Mod. Phys. {\bf 46}, 465 (1974).

\bibitem{Herman17a} F. Herman and R. Hlubina, Phys. Rev. B {\bf 95}, 094514 (2017).  

\bibitem{Herman18} F. Herman and R. Hlubina, Phys. Rev. B {\bf 97}, 014517 (2018).

\bibitem{Anderson59} P. W. Anderson, J. Phys. Chem. Solids {\bf 11}, 26 (1959).

\bibitem{DeGennes66} P. G. de Gennes, {\it Superconductivity of Metals and Alloys} (Benjamin, New York, 1966).

\bibitem{note} Strictly speaking, the mean free path is given by $\ell\equiv \hbar v_F/(2\Gamma_{\rm tot})$, where $\Gamma_{\rm tot}=\Gamma_s+\Gamma$. However, usually $\Gamma_s\gg \Gamma$.

\bibitem{limit} In the asymptotic limit of very large $\Gamma_s$, we find $\Lambda=\sqrt{2}\Gamma_s$ correspoding to $b\approx 0.71$. However, the asymptotic limit is reached for unphysically large values of $\Gamma_s$.

\bibitem{note2}But, of course, outside the fluctuation-dominated region in the immediate vicinity of $T_c$.

\bibitem{Rickayzen80} G. Rickayzen, {\it Green's Functions and Condensed Matter} (Academic, New York, 1980).

\end{thebibliography}
\end{document}